\documentclass[12pt]{iopart}

\usepackage{iopams}
\usepackage[latin1]{inputenc}
\usepackage{setstack}
\usepackage{graphicx}
\usepackage[capitalise]{cleveref}

\usepackage[normalem]{ulem}

\begin{document}

\title{Versatile electric fields for the manipulation of ultracold NaK molecules}

\author{M.W.~Gempel, T.~Hartmann, T.A.~Schulze, K.K.~Voges,
 A.~Zenesini and S.~Ospelkaus}

\address {Institut f\"{u}r Quantenoptik \& Laboratorium für Nano- und Quantenengineering, Leibniz Universit\"{a}t Hannover, 30167 Hannover, Germany}
\ead{silke.ospelkaus@iqo.uni-hannover.de}

\begin{abstract}
	In this paper, we present an electrode geometry for the manipulation of ultracold rovibrational ground state NaK molecules. The electrode system allows to induce a dipole moment in trapped diatomic NaK molecules with a magnitude up to $68 \%$ of their internal dipole moment along any direction in a given two-dimensional plane. The strength, the sign and the direction of the induced dipole moment is therefore fully tunable. Furthermore, the possibility to create strong electric field gradients provides the opportunity to address molecules in single layers of an optical lattice.
	The maximal relative variation of the electric field over the trapping volume is below $10^{-6}$. At the desired electric field value of 10 kV/cm this corresponds to a deviation of 0.01 V/cm.
	The electrode structure is made of transparent indium tin oxide and combines large optical access for sophisticated optical dipole traps and optical lattice configurations with the possibility to create versatile electric field configurations. 
\end{abstract}

\section{Introduction}

	Atomtronics heads for the development of novel technological devices based on quantum systems of cold and ultracold atoms and is expected to have an impact on future technology comparable to the invention of solid-state transistors in the 60'. Recent remarkable progress in this young field includes the realization of atomic circuits implemented either with planar electronics circuits, so-called `atom-chips' \cite{kru2003}, or by `painting' light potentials on demand \cite{hen2009}. The different pathways for the realization of atomtronics devices are object of intense research and continuous progress and are discussed in this Special Issue. 
	
	While the name \textit{atomtronics} suggests that these future devices will be realized based on atomic quantum objects, the use of molecular quantum systems \cite{jin2012} with unique properties might open additional opportunities. 
	The field of ultracold quantum gases of molecules has recently seen tremendous progress opening new prospects for fundamental research and technological applications. Polar molecules promise to be an excellent test ground for fundamental laws of nature \cite{acm2014}, few and many-body phenomena \cite{bar2012} and novel quantum computing schemes \cite{dem2002}. 
	Ensembles of ultracold, polar, diatomic molecules are considered to be one of the most promising candidates for the investigation of strongly correlated quantum many-body systems due to strong long-range and anisotropic dipole-dipole interactions \cite{bar2012, ni2008, yan2013}.
	
	The wealth of opportunities with ultracold molecules is due to the complex molecular structure with rovibrational quantum degrees of freedom, which, at the same time, represents a severe hurdle in the implementation of efficient cooling, preparation and control schemes for ultracold molecular quantum objects. Several extended review papers have been recently published focusing in particular on the issue of molecular cooling and we refer to them for a detailed discussion of different methods and techniques \cite{hut2012, que2012, van2012, car2009}.	 
	  
	Another very demanding aspect in the control of ultracold molecules is the generation of stable and highly controllable electric fields on the order of $10\,\mathrm{kV/cm}$ to induce and control the molecular dipole moment. 
	The flexible and precise tuning of strength and angle of the induced dipole moment is a stringent requirement that, when achieved, will open the way for the realization of exciting theoretical proposal on molecular dynamics and novel quantum phases as for example \cite{Tic2011, Fev2014, Que2015, Zha2015}.
	Several groups currently work to obtain ultracold samples of polar molecules.
	Some of them were already able to obtain ground-state molecules (KRb \cite{ni2008}, LiCs \cite{Deiglmayr2008}, RbCs \cite{Molony2014,Takekoshi2014}, NaK \cite{Park2015}, NaRb \cite{Guo2016}).  
	For all of these groups the control of the electric field, polarizing the molecules, will be an important technical topic to solve.
	
	In this paper, we discuss the design and realization of an electrode system for the manipulation and control of ultracold NaK molecules. 
	Ultracold samples of polar, rovibrational ground state NaK molecules will be prepared in our experimental apparatus through the association of ultracold Na and K atoms \cite{wu2012,sch2013} and are potential candidates for molecular dipolar quantum objects in future technological \textit{atomtronics} devices. 

	The presented electrode system meets the requirements regarding versatility, stability and large optical access mentioned above.
	It allows to apply strong, tunable, homogeneous electric fields, of up to $10\,\mathrm{kV/cm}$, along any direction in a given two-dimensional plane. 
	The strength and the direction of the induced dipole moment is therefore fully tunable.
	At $10\,\mathrm{kV/cm}$ the induced dipole moment of NaK ground-state molecules corresponds to $68 \%$ of their internal dipole moment. 
	In addition the electrode system can be use to create strong electric field gradients, providing the opportunity to address molecules in single layers of an optical lattice.
	
	In section 2 we review the principle of the induced dipole moment in the rigid rotor model and present some estimates on the experimental requirements. In section 3, we discuss the critical influence of inhomogeneous electric fields on trapped molecules discussing in particular field gradients and curvatures. Section 4 reviews our numerical simulations on different electrode geometries and describes in detail our chosen experimental system. Finally, we summarize our findings and outlook the possibility to extend our experimental system to the manipulation of molecules in on-chip systems.

\section{Molecules in Electric Fields}
	\label{MolecularDipoles}
	
	In this section, we review the basic formalisms to describe the effect of electric fields on so-called rigid rotor molecules such as bi-alkali molecules \cite{boh2010}.
	Neglecting e.g. the electric quadrupole effect, we assume that the rigid rotor fully describes the rotational structure of the molecules in the vibrational ground state. The Hamiltonian $H^{0}$ of the molecule is given by:
	\begin{equation} \label{Eq:Hamilton0}
		H^0_{ij} \sim B \cdot N_i\left( N_i+1 \right)\delta_{ij} ~\mathrm{.}
	\end{equation}
	Here, $H^0_{ij}$ is given in the basis of its eigenfunctions $\left| N_i,m_{N_i} \right\rangle $, which are the spherical harmonics. $B$ is the rotational constant of the molecule and $\delta_{ij}$ is the Kronecker delta. 
	The quantum numbers specify the rotation $N_i$ and its projection $m_{N_i}$.
	As $\left| N_i,m_{N_i}\right\rangle$ are parity eigenstates, a molecule prepared in a specific rovibrational quantum state $\left|N_i,m_{N_i}\right\rangle $ cannot have an electric dipole moment in a space-fixed reference frame. 
	
	The molecular dipole moment only manifests in a space-fixed frame when applying an external field, which breaks the spatial isotropy of $H^{0}$.
	The coupling between an applied electric DC field $\vec{\mathcal{E}}$ and the molecule's internal dipole moment $\vec{d}$ mixes $\left|N_i,m_{N_i}\right\rangle $ of different parity.
	The new eigenstates are associated with a finite dipole moment in the laboratory frame and are obtained by diagonalizing the Hamiltonian
	\begin{equation}
			H_{ij} \sim 
			B \cdot \left[ N_{i}\left( N_{i}+1 \right) \right] \delta_{ij}
			- ~\!\left\langle N_i,m_{N_i}\right| \vec{\mathcal{E}} \cdot \vec{d} \left|N_j,m_{N_j}\right\rangle \, \mathrm{.}
	\label{Eq:Hamilton1}
	\end{equation}
	Aligning the electric field along the vertical coordinate z, $\vec{\mathcal{E}}$ and $\vec{d}$ enclose the polar angle $\theta$.
	
	Rewriting $H_{ij}$ in dimensionless quantities leads to:
	\begin{equation}
			H_{ij} \sim 
			\left[ N_{i}\left( N_{i}+1 \right) \right] \delta_{ij}
			- E\left\langle N_i,m_{N_i}\right| \cos\theta \left| N_j,m_{N_j}\right\rangle\, \mathrm{.}
	\label{Eq:Hamilton}
	\end{equation}
	In this representation energies are in units of $B$ and electric fields in units of $B/d$. 
	The dimensionless electric field $E$ thus gives a direct measure of the coupling between rotational states.
	For NaK, the dipole moment of the vibrational ground state manifold is $d= 2.76 \, \mathrm{Debye}$ \cite{wal2013} and the rotational constant $B$ is $h\times 2.83 \,\mathrm{GHz}$, which results in a characteristic electric field of $B/d = 2.04 \, \mathrm{kV/cm}$. 
	Here, $h$ is the Planck constant.

	\begin{figure}[t]
	\centering
		\includegraphics[width=10cm]{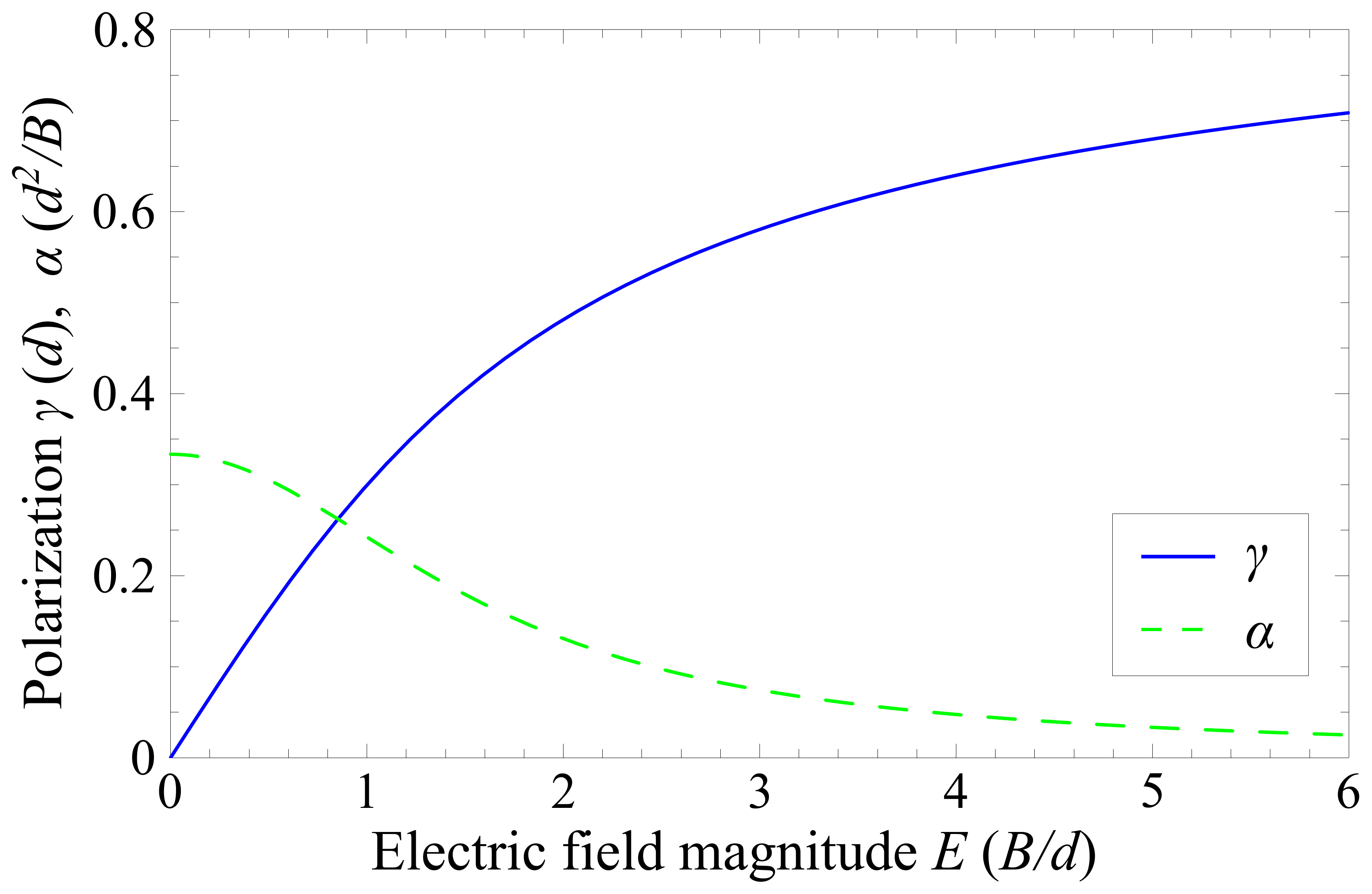}
		\caption{Induced dipole moment and its first derivative. Solid line: Induced dipole moment $\gamma$ as a function of $E$. Dashed line: $\alpha$, is the second derivative of $\left| U_{A}\right| $. The molecular energy follows a quadratic and a linear Stark effect in the limit of low and high electric field, respectively.}
		\label{fig:DipoleMomentSaturation}
	\end{figure}

	The new ground state of the molecule $\left|A\right>$ converges to $\left|0,0\right>$ for $E\to 0$ and its energy is given by $U_\mathrm{A}(E)=\left<A\right|H\left|A\right>$.
	The induced dipole moment of $\left|A\right>$, $\gamma$, lines up with the quantization axis $z$, which is defined by the direction of $\vec{\mathcal{E}}$, and is given by the first derivative of $U_\mathrm{A}$ with respect to $E$:
	\begin{equation}
		\gamma = - \frac{\partial U_{\mathrm{A}}}{\partial E} ~\mathrm{.}
	\end{equation}
	In Fig.~\ref{fig:DipoleMomentSaturation} we plot the induced dipole moment $\gamma$, in units of $d$, as a function of $E$, up to $6.0\,B/d$.
	The quantity $\gamma$ rises linearly for small $E$, but saturates at a value of one for $E \to \infty$ where the electric field dominates all energies in Eq.~\ref{Eq:Hamilton}.
	In the intermediate region the coupling of $\left| 0,0\right\rangle $ to higher rotational states provides a smooth transition between both regimes.
	At $E = 5.0\,\left( B/d\right)$, which corresponds to $10\,\mathrm{kV/cm}$ in the case of a NaK ground-state molecule, an induced dipole moment of $0.68\,\left( d\right)$, corresponding to approximately $1.9\,\mathrm{Debye}$, is reached.
	When giving numeric values for the dimensionless variables, like `$E = 5.0\,\left( B/d\right)$', the corresponding conversion to SI units is indicated by brackets. 
	Numeric values of the electric field given in SI units (like `$10\,\mathrm{kV/cm}$') always refer to NaK.
	
	The next section will focus on the effects of spatial variations in $E$ on the trapping of a molecule.

\section{The Effect of Electric Fields on Trapped Molecules}
\label{Sec:ElectricFields}

\begin{figure}[b]
		\centering
		\includegraphics[width=10cm]{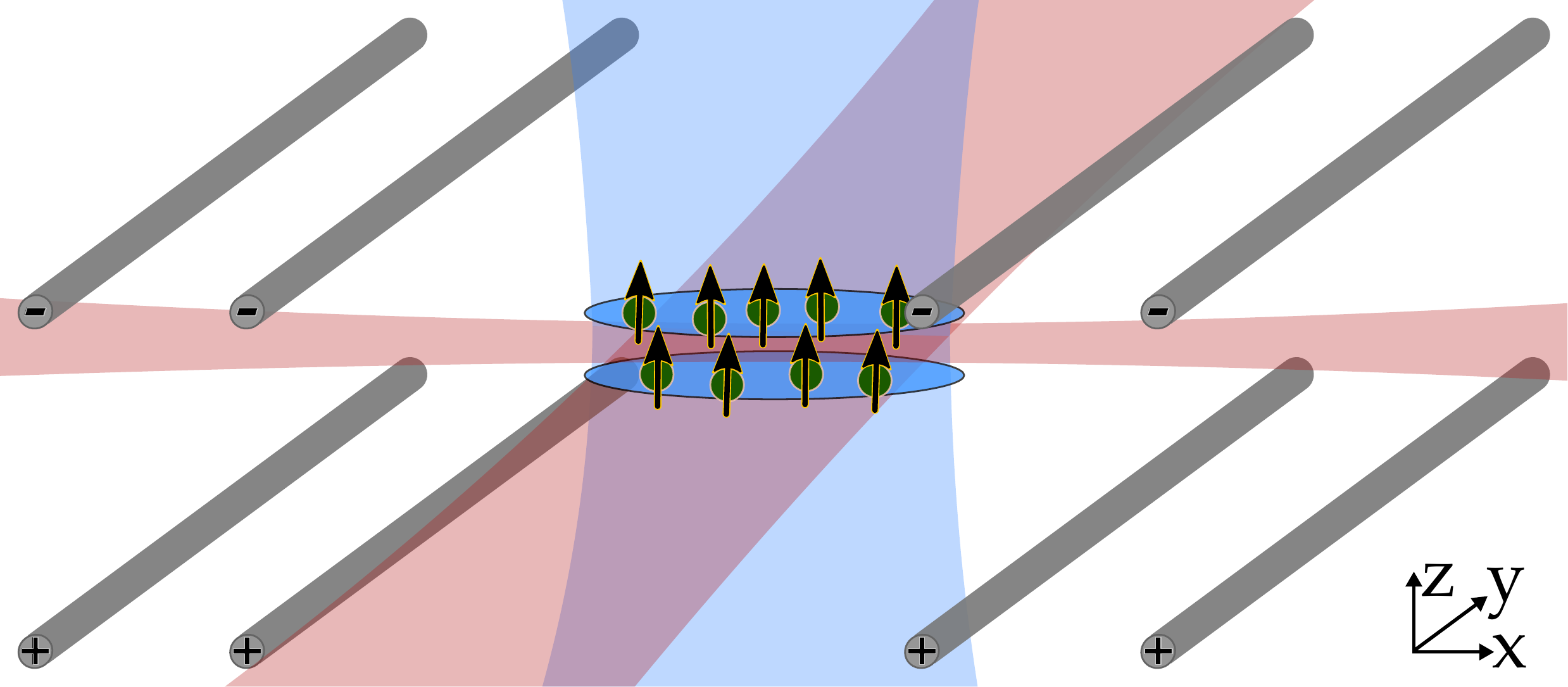}
		\caption{
		Three-dimensional view of the horizontally crossed optical dipole trap with one-dimensional optical lattice in the vertical direction.
		The two dipole trap beams, propagating in the x and y direction are sketched in red, the vertical optical lattice in blue.
		Two layers of the optical lattice are indicated.
		The gray rods indicate the electrodes extending in the y direction, which create the electric fields in the x-z plane to polarize the molecules.
		The arrows indicate the induced dipole moments of the molecules.}
		\label{fig:ExpSetup}
	\end{figure}

	Due to the Stark shift, the energy of a ground-state molecule is strongly affected by temporal and spatial changes of $E$, $U_{\mathrm{A}} = U_{\mathrm{A}} \left( E \right)$.
	These changes can either create excitations or deform the overall spatial potential $U_{\mathrm{tot}}$ of a molecule, which can even lead to a loss of molecules from the trap. 
	In our experiment the overall spatial confinement of NaK is obtained by two red detuned, far off-resonant, horizontal, crossed optical dipole trap beams, see Fig.~\ref{fig:ExpSetup}. In the same figure, the electrodes are depicted as rods extending in the y direction.
	The direction of the electric field resulting from the applied potentials can be inferred by the directions of the induced dipole moments of the molecules (arrows in picture). 
	A vertical one-dimensional optical lattice can be added to the trap.
	It creates a stack of layers in which the molecules are confined in a quasi two-dimensional potential.
	Further lattice beams along the directions of the optical dipole trap can imprint a full three-dimensional lattice (not depicted here).
	
	When being exposed to an applied electric field, an optically trapped polar molecule experiences a position dependent total energy shift given by 
	\begin{equation}
		U_{\mathrm{tot}} = U_{\mathrm{dip}} + U_{\mathrm{el}} ~\mathrm{.}
	\end{equation}
	where $U_\mathrm{dip}$ and  $U_{\mathrm{el}}$ are the energy shifts from the confining potential and the DC Stark shift respectively. 
	For simplicity we limit the following considerations regarding $U_{\mathrm{tot}}$ to the horizontal direction $x$. 
	The treatment in the other two directions is analogous.
	The confinement, provided by the optical dipole trap, can be parametrized as a Gaussian profile (see Fig.\ref{fig:DTDeform}\,A)
	\begin{equation}
		U_{\mathrm{dip}} = 	-U_{\mathrm{dip,0}} e^{-2 \left( x/w_{0} \right)^2} =
					-U_{\mathrm{dip,0}} e^{-2 \tilde{x}^2} 
	\end{equation}
	with a maximum trap depth $U_{\mathrm{dip,0}}$, in units of $B$.
	The spatial dimension $\tilde{x} = \frac{x}{w_{0}}$ is normalized to the waist $w_{0}$ of the optical dipole trap, which is $100~\mathrm{\mu m}$ in our setup.
	For a molecule in its rovibrational ground state the potential created by the electric field is given by
	\begin{equation}
		U_{\mathrm{el}} = U_{\mathrm{A}} \left( E \left( \tilde{x} \right) \right) 
		\mathrm{.}
	\end{equation}
	Therefore, spatial variations $\delta E$ of the electric field lead to perturbations of the molecules' confinement.
	In the worst case the change of the electric field over the optical dipole trap can lead to deconfinement of the molecules. 
	As such perturbations are most severe at high polarization, i.e. electric field, we assume $U_{\mathrm{el}}= \left. U_{\mathrm{A}} \right|_{E_{0}} 
			- \left. \gamma \right|_{E_{0}}
			 \delta E + \mathcal{O} \! \left(\delta E^2\right)$.

\subsection{Different Contributions of the Electric Field}

\begin{figure}[t]
		\centering
		\includegraphics[width=10cm]{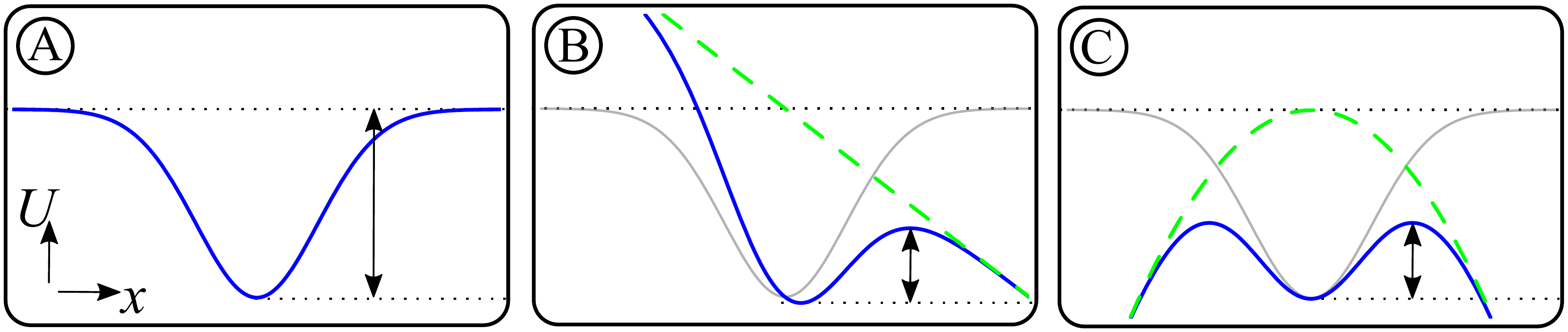}
		\caption{
		Deformation of $U_{\mathrm{tot}}$ due to electric fields.
		(A) shows the cross section of an unperturbed optical dipole trap potential $U_{\mathrm{dip}}$. (B) visualizes the effect of an additional potential gradient along the $x$ direction, (C) shows the effect of an electric field with curvature on the confining potential.
		The arrows indicate the trapping depth of the resulting potential.}
		\label{fig:DTDeform}
	\end{figure}

	The influence of the electric field on the trapping of the molecules can be most easily described when expanding the spatial dependence of $E(\tilde{x})$ in a Taylor series of $\tilde{x}$ around the symmetry center of the electrode geometry at $\tilde{x}=0$:
	\begin{equation} \label{Eq:Taylor}
		E\left( \tilde{x} \right) = E_{0} + E_{1} \tilde{x} + E_{2}\tilde{x}^2 + \mathcal{O} \! \left( \tilde{x}^3 \right), 
	\end{equation}
	where $\left.E_1=\frac{\partial E}{\partial\tilde{x}}\right|_{\tilde{x}=0}$ and $\left.E_2=\frac{\partial^2 E}{2\partial\tilde{x}^2}\right|_{\tilde{x}=0}$ are the gradient and curvature of the electric field, respectively. For electrode geometries, whose characteristic length scale $l$ is much larger than $w_{0}$, truncating $\mathcal{O} \! \left( \tilde{x}^3 \right)$ is a good approximation, as terms of the order $\tilde{x}^n$ scale with $\left(\frac{w_{0}}{l}\right)^n$.
	In our setup, the electrode geometry has a characteristic length scale of 10 mm ( $\frac{w_{0}}{l} \approx 0.01$) and terms of higher orders in $\tilde{x}$ are naturally suppressed.
	By using $\tilde{x}$ instead of $x$, the gradient $E_{1}$ and curvature $E_{2}$ correspond to the absolute field change over the length scale of the optical dipole trap.
	Hence $E_{1}$ and $E_{2}$ can be compared directly to the spatially uniform term $E_{0}$. 
	To obtain the actual gradient and curvature, $E_{1}$ and $E_{2}$ must be divided by $w_{0}$ and $w_{0}^2$ respectively.	
	The total potential, with electric field terms up to the second order in the spatial deviation in the electric field is given by
		\begin{eqnarray} 
	\label{Eq:Taylor2}
		\fl U_{\mathrm{tot}} 
		& = -U_{\mathrm{dip,0}} e^{-2 \tilde{x}^2}
		+ U_{\mathrm{A}}\left( E_{0} + \left[ E_{1} \tilde{x} + E_{2} \tilde{x}^2 \right] + ... \right) \\
		\fl & \approx -U_{\mathrm{dip,0}} e^{-2 \tilde{x}^2} 
		+ \left. U_{\mathrm{A}} \right|_{E_{0}} 
		- \left. \gamma \right|_{E_{0}}
		 \left[ E_{1} \tilde{x} + E_{2} \tilde{x}^2 \right]
		-\mathcal{O} \! \left( \left[ E_{1} \tilde{x} + E_{2} \tilde{x}^2 \right]^2 \right).
		\end{eqnarray}
	 As illustrated in Fig.~\ref{fig:DTDeform}\,B, the gradient term $E_{1}$ tilts the confining potential $U_{\mathrm{tot}}$.	 
	The curvature of the electric field $E_{2}$ reduces the trap depth and the trap frequency in at least one spatial direction \cite{ear1842} as can be seen in Fig.~\ref{fig:DTDeform}\,C.
	In the following sections, we will discuss these effects quantitatively and we will deduce requirements on the electrode system. 
	
	\subsubsection{Homogeneous fields }

	The constant term $E_{0}$ does not change the shape of $U_{\mathrm{tot}}$. However, when considering temporal variations of the electric field, it can lead to excitations of molecules to higher rotational states. But excitations will only occur when \cite{von1929}
	\begin{equation}
		\frac{\partial E_{0}}{\partial t} \gg 12 \frac{B}{h} ~\mathrm{.}
	\end{equation}
	This corresponds to an electric field noise of $100\,\mathrm{kV/cm}\times\mathrm{GHz}$ or a change in the electric field of $10 ~\mathrm{kV/cm}$ in $0.1 ~\mathrm{ns}$ for NaK - a ramping speed far from being experimentally relevant. 
	
	\subsubsection{Higher orders in $x$}

	The overall potential including the effect of an electric field gradient is 
	\begin{eqnarray}
		U_{\mathrm{tot}}
		\approx -U_{\mathrm{dip,0}} e^{-2 \tilde{x}^2} 
		- \left. \gamma \right|_{E_{0}} \left[ E_{1} \tilde{x} \right] 
		\mathrm{.}
	\end{eqnarray}
		The minimum of such a potential, shown in Fig.~\ref{fig:DTDeform}\,B, exists only for an electric field gradient $E_{1}$ smaller than:
	\begin{eqnarray}
		E_{1,\mathrm{lim}} = \frac{2}{\sqrt{e}} \frac{U_{\mathrm{dip,0}}}{\left. \gamma \right|_{E_{0}}} ~ \mathrm{.}
		\label{Eq:GradLim}
	\end{eqnarray}
	
	This limiting value disregards losses due to tunneling across the barrier at the open side of the tilted potential, which is indicated by the black arrow in Fig.~\ref{fig:DTDeform}\,B.
	Those tunneling events become, however, only relevant for $E_{1} \approx E_{1,\mathrm{lim}}$ and can be neglected for the derivation of $E_{1,\mathrm{lim}}$. 
		
	The effect of the electric field curvature $E_{2}$ (see Fig.~\ref{Eq:Taylor2}) on the potential is illustrated in Fig.~\ref{fig:DTDeform}\,C and gives
	\begin{eqnarray}
			U_{\mathrm{tot}} 
			\approx -U_{\mathrm{dip,0}} e^{-2 \tilde{x}^2}
			- \left. \gamma \right|_{E_{0}} \left[ E_{2} \tilde{x}^2 \right] 
			\, \mathrm{.}
	\end{eqnarray}
		In this case the trapping is only possible for curvatures smaller than
	\begin{eqnarray}
		E_{2,\mathrm{lim}} = \frac{2 U_{\mathrm{dip,0}}}{\left. \gamma \right|_{E_{0}}} \, \mathrm{.}
		\label{CurvLim}
	\end{eqnarray}
	
	Table~\ref{Tab:Limits} summarizes our findings for $E_{1 ,\mathrm{lim}}$ and $E_{2 ,\mathrm{lim}}$ and gives typical values for the NaK molecules. We consider a dipole trap depth of $U_{\mathrm{dip,0}} \cdot B = k_{B} \times 10\mathrm{µK}$ and an induced dipole moment of 68 \% of the internal dipole moment. In this case the spatial variation of the field across the dimension of the optical dipole trap given by the $w_0$ has to be lower than $10^{-4} E_0$ that means four orders of magnitude lower than the field required to polarize. 
	
	When using an optical lattice along the direction of the gradient (or curvature) the lattice spacing takes the role of $w_{0}$. In the case of a retro-reflected $\lambda = 1064~\mathrm{nm}$ lattice the spacing is $532~\mathrm{nm}$ and orders of magnitude smaller than $w_{0}$. 
	Accordingly, losses due to the deformation and the resulting `opening' of the trapping potential, as shown in Fig.~\ref{fig:DTDeform} for the optical dipole trap, can be strongly suppressed by applying an optical lattice.
	Here we have focused our attention on the constraint to maintain a highly constant electric field. In the same manner one can consider the use of electric field gradients to investigate new physics, similar to experiments performed in the recent years in tilted optical lattices, e.g in \cite{sia2007, hal2010, sim2011}
	
	\begin{table}[t] 
	\begin{center} 
	\begin{tabular}{c|c|c} 
		Coefficient & Limit & $\mathrm{NaK_{68\%}}$ \\ 
	\hline \rule[-2ex]{0pt}{5.5ex} $E_{1,\mathrm{lim}}$ & 
		$ \frac{2}{\sqrt{e}} \frac{U_{\mathrm{dip,0}}}{\gamma} $
		 &
		$2.7 \cdot 10^{-4} ~ \mathrm{\frac{kV}{cm}}\frac{1}{w_{0}}$ \\ 
	\hline \rule[-2ex]{0pt}{5.5ex} $E_{2,\mathrm{lim}}$ & 
		$ 2 \frac{ U_{\mathrm{dip,0}}}{\gamma}$ 
	 &
		$4.5 \cdot 10^{-4} ~ \mathrm{\frac{kV}{cm}}\frac{1}{w_{0}^2}$ \\ 
	\end{tabular} 
	\caption{ Column 1 and 2: $E_{1,\mathrm{lim}}$ and $E_{2,\mathrm{lim}}$ as given in the text. Column 3: Experimentally relevant example for NaK as explained in the text. }
	\label{Tab:Limits}
	\end{center}
	\end{table}
	
	Until now we assumed the symmetry center of the electric field to coincide with the center of the optical dipole trap. In a system with electric field curvature, it is however necessary to also consider a displacement $\tilde{x}_0$ between the electric field and the dipole trap center. In this case, the total confining potential is given by:
	\begin{eqnarray}
			U_{\mathrm{tot}} 
			\approx -U_{\mathrm{dip,0}} e^{-2 \tilde{x}^2}
			- \left. \gamma \right|_{E_{0}} \left[ E_{2,x_0} \left( \tilde{x}-\tilde{x}_0 \right) ^{2} \right] 
	\end{eqnarray}

	In the case of $\tilde{x}_0 \to 0$, we recover the curvature limit found above. 
	However, for finite $\tilde{x}_0$, $E_{2,x_0,\mathrm{lim}}$ can be approximated	\footnote{The given formula is an approximation. It deviates by at most $20 \%$ from the correct value given by 
				$E_{2,x_0 ,\mathrm{lim}} = 
					\frac{2 U_{\mathrm{dip,0}}}{\left. \gamma \right|_{E_{0}}} \left(\frac{\eta^2}{9} - 1\right)
					 e^{ -\frac{1}{18} \eta^2} 
				\mathrm{, ~~with~~}
				\eta = \zeta + \frac{4 \tilde{x}_0^2}{\zeta}-2 \tilde{x}_0
				\mathrm{~~and~~}
				\zeta = \sqrt[3]{-8 \tilde{x}_0^3+3 \sqrt{81 \tilde{x}_0^2-48 \tilde{x}_0^4}+27 \tilde{x}_0}
			$.} by
	\begin{equation}
		E_{2,x_0,\mathrm{lim}} = \frac{U_{\mathrm{dip,0}}}{\left. \gamma \right|_{E_{0}}} 
		\frac{1}{\sqrt{e} \cdot \tilde{x}_0 + 0.5} ~\mathrm{.}
	\end{equation}

	\subsection{Addressing single layers using electric field gradients}
	\label{ElectricalFieldGradientsInOpticalLattices}
	
	In our experiments, we aim to selectively address molecules in single layers of a stacked one-dimensional optical lattice as the one shown in Fig.~\ref{fig:ExpSetup}. 
	This can be realized by applying a strong electric field gradient along the lattice. 
	The Stark shift then introduces a layer dependent transition energy for the excitation from the ground state $\left|A\right>$ to the first rotationally excited state $\left|B\right>$, which goes to $\left|1,0\right>$ for $E \to 0$. By choosing a specific microwave frequency, it is thus possible to selectively drive the transition in a specific layer, similarly to what has been demonstrated for atoms by means of magnetic field gradients \cite{wei2011}.

		\begin{figure}[t]
				\centering
				\includegraphics[width=10cm]{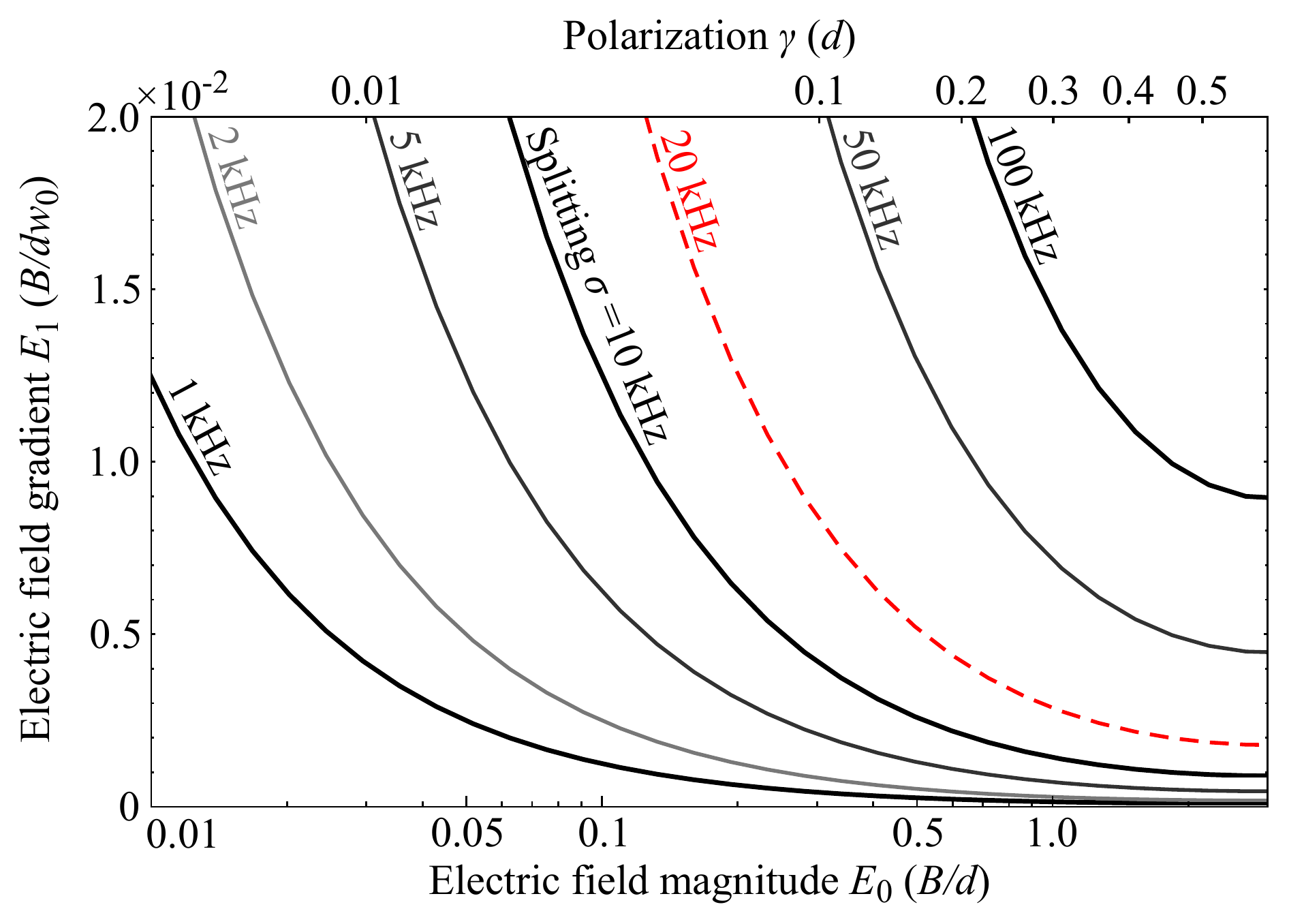}
				\caption{
					Contours of constant splitting $\sigma$ of the microwave driven transition from $\left| A\right\rangle$ to $\left| B\right\rangle$, in NaK.
					The contours indicate the splitting $\sigma$ of the transition from $\left| A\right\rangle$ to $\left| B\right\rangle$ between two layers of an optical lattice with $532 ~\mathrm{nm}$ spacing, when applying a certain electric field gradient $E_{\mathrm{1}}$ (y-axis) and background field $E_{\mathrm{0}}$ (x-axis).
					A splitting of $\sigma = 20~\mathrm{kHz}$, corresponding to a typical bandwidth of a microwave pulse, is indicated by the dashed, red line.
					The upper x-axis gives the polarization $\gamma$ corresponding to the electric field given on the lower x-axis.
				}
				\label{fig:FieldAndGradientForSplitting}
			\end{figure} 

	This can e.g. be useful in a bilayer system, when the molecules of one layer have to be prepared in a different rotational state than the molecules in the other layer, a situation envisioned e.g. in \cite{pik2011}.
	When, however, considering a microwave pulse with a finite bandwidth $\sigma'$, the transition can only be selectively driven in one of the layers when it is energetically detuned by much more than $\sigma'$ in the neighboring layer.
	In our experiment, with a layer spacing of $532 ~\mathrm{nm}$, we typically choose a bandwidth of $\sigma'=20\,\mathrm{kHz}$. 
	To selectively address one of the layers, will require an applied gradient much greater than $0.97\,\left( B/dw_{0}\right)$ when $E_{\mathrm{0}}=0$.
	In the case of NaK, this corresponds to $2.0 \cdot 10^2 ~\mathrm{kV/cm^2}$, a gradient which is very hard to achieve in our setup.
			
	However, as shown in Fig.~\ref{fig:FieldAndGradientForSplitting} the splitting of the transition between different layers not only increases with the electric field gradient $E_{\mathrm{1}}$ (y-axis), but also with an homogeneous background field $E_{\mathrm{0}}$ (x-axis), because the the molecules enter the regime of the linear Stark effect \footnote{
	At a certain $E_{\mathrm{0}}$ this increase of the splitting reverses as $\left<B\right| H \left|B\right>$ is repelled by energetically higher states}.
	At $E_{0} = 0.50\,\left( B/d\right)$ the gradient required for a splitting of $20\,\mathrm{kHz}$ has already reduced to $5.2 \cdot 10^{-3}\,\left( B/dw_{0}\right)$. 
	In the case of NaK this corresponds to $1.1~\mathrm{kV/cm^2}$, which is a realistic value for experimental realization (see Sec.\,\ref{MolecularDipoles}).

\section{Design of an electrode configuration for the creation of versatile electric fields} 
\label{PossibleElectrodeconfiguration}

\begin{figure}[t]
	\centering
	\includegraphics[width=12cm]{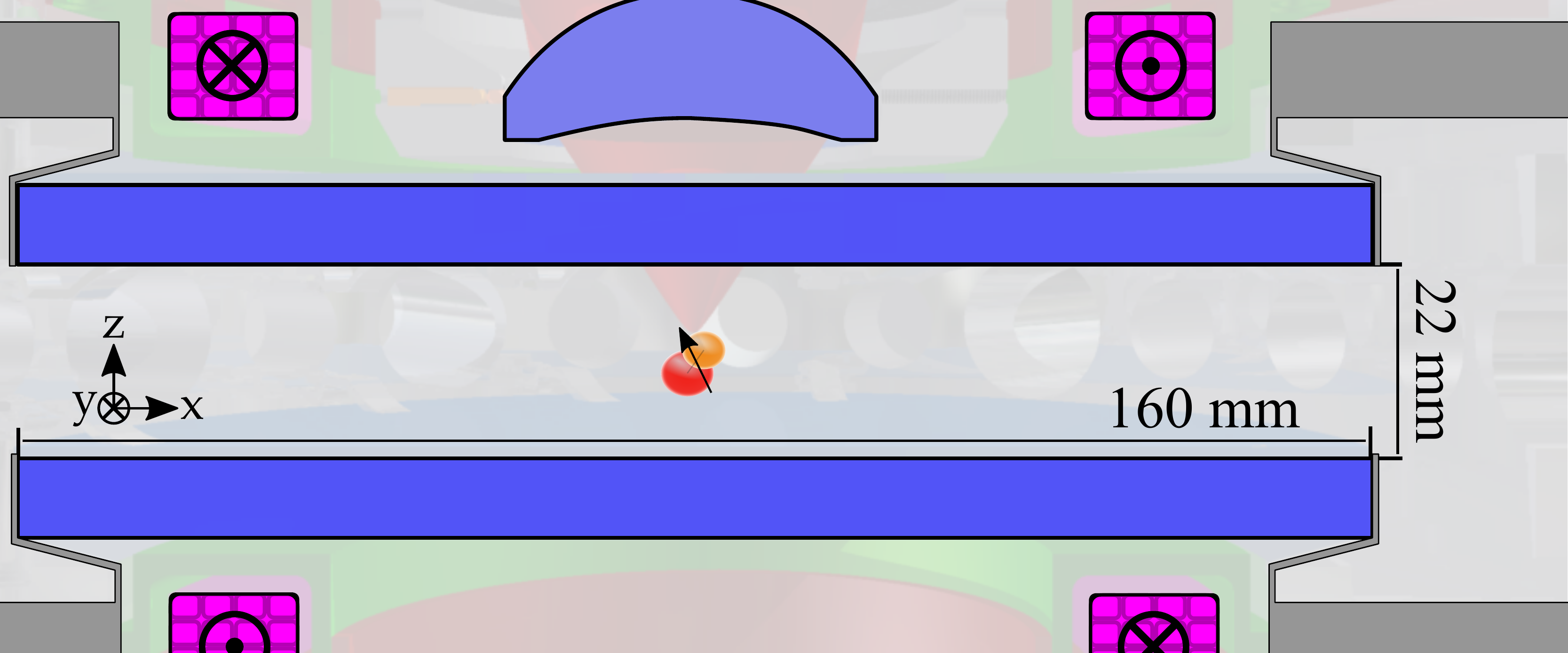}
		\caption{
		Cross section of the vacuum chamber with coils and high resolution imaging lens.
		The solenoids creating the the magnetic fields are shown in pink.
		The gray structures indicate the flanges in which the vacuum windows are mounted. 
		The two main windows themselves are shown in blue. 
		They are spaced by $22 ~\mathrm{mm}$.
		In the center of the upper solenoid the first lens of the high resolution imaging system is depicted.
		The sketch of a polar molecule in the center indicates the location of the trapping region.}
	\label{fig:ExpSit}
	\end{figure}

	After outlining the requirements on the electric field, the next step is the design of the electrode geometry. Figure \ref{fig:ExpSit} shows a cross section of our vacuum chamber for the creation and manipulation of polar molecules. 
	In the center of the chamber (indicated by the sketch of a polar molecule) all preparation steps for molecule creation take place. This includes magneto-optical trapping of sodium and potassium, trapping of the two atomic species in a magnetic trap and evaporative cooling to quantum degeneracy, magneto-association of the two atomic species to Feshbach molecules and subsequent transfer of the molecules to the rovibrational ground state. 
	To provide enough space for obtaining high atom numbers in the magneto-optical trap (MOT) we separated the two opposing main windows of the chamber by $22~\mathrm{mm}$. 
	The windows have a large diameter of $160~\mathrm{mm}$, which provides enough space to recess the solenoids creating magnetic fields and provides large optical access for high resolution imaging.
	
	To preserve the optical access, the electrodes creating the external electric fields for manipulating molecules are implemented as an indium tin oxide (ITO) coating, of $167~\mathrm{nm}$ thickness, on the two opposing main windows.
	ITO is electrically conducting and at the same time transparent in a broad range of the visible spectrum.
	The coating is deposited on the vacuum side of the the windows to reduce charged particle deposition on the glass surfaces.
	The measured transmission of a single window at the sodium D2 line ($589~\mathrm{nm}$) is $\mathrm{>99\%}$ in the regions without ITO and $\mathrm{>92\%}$ in the regions with ITO.
	As the transmission of ITO in the infrared is less favorable, the coating must have a gap in the center where infrared vertical beams, e.g. for an optical lattice, are supposed to pass. 
	This gap provides a transmission of more than $\mathrm{99.9\%}$ at a wavelength of $1064 ~\mathrm{nm}$.
	
	Coating the electrode structure on the main windows allows for arbitrary electrode configurations in the two-dimensional plane given by the window surface. However, in the experiment, we aim for the simplest possible electrode configuration fulfilling our requirements on the creation of homogeneous fields and gradients, and on the compensation of curvature as discussed above. In the following, we therefore start our discussion with parallel plate capacitors, then extend the discussion to a four-rod-like system and finally present our eight-rod-like geometry. 

	\subsection{Plate capacitors}
	\label{PlateCapacitor} 

	A system of parallel plate capacitors is the simplest possible geometry to consider. It has the advantage to create very uniform fields at its geometric center	but the direction of the electric field can only be changed by either physically turning the plates, or by adding a second parallel plate capacitor.	In principle, two orthogonal capacitors can then create electric fields along any direction within a two-dimensional plane.
	Considering two orthogonal capacitors with a plate spacing of $22~\mathrm{mm}$ the dimensions of a plate in our setup would be limited to $136 \times 15 ~ \mathrm{mm^2}$, to avoid discharging.
	The curvature to magnitude ratio resulting from this geometry is $E_{2}/E_{0} = 2.4 \cdot 10^{-5}$.
	At field amplitudes of $E_{0} = 10~\mathrm{kV/cm}$ this already reaches the order of magnitude of the critical value of the curvature derived in Tab.~\ref{Tab:Limits}.

	\subsection{Four rods}
	\label{FourRods}

\begin{figure}[h]
		\centering
		\includegraphics[width=16cm]{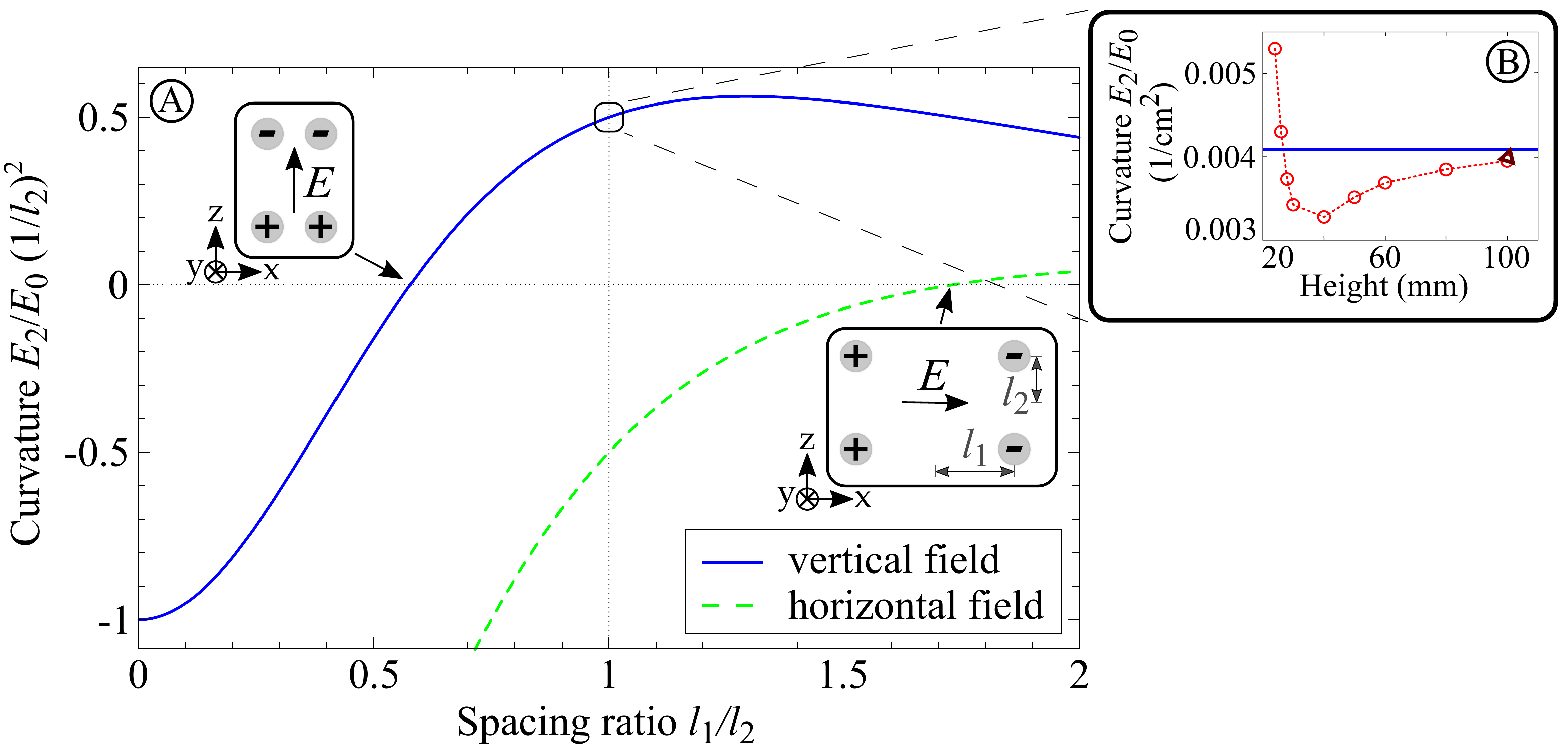}
		\caption{
		(A) $E_{2}/E_{0}$ of four infinite rods as a function of the electrode spacing ratio $l_{1}/l_{2}$.
		The two insets show the arrangement of the four infinitely long rods, as a cross sections through the x-z plane. 
		The horizontal and vertical spacing for the electrodes are $2l_{1}$ and $2l_{2}$.
		The signs of the electric potentials, necessary to obtain the different field directions, are indicated as '+' or '-'.
		The insets depict situations for which the curvature of a vertical or horizontal electric field vanishes, at $l_{1}/l_{2} = 0.58$ and $l_{1}/l_{2} = 1.7$.
		The blue line in the detail view (B) on the right is the analytic solution of $E_{2}/E_{0}$ for $l_{1} = l_{2} = 11 \,\mathrm{mm}$.
		The red circles are the results of COMSOL simulations, plotted as a function of the height of the volume which was used for the simulation.
		The diameter of the volume was $160 \,\mathrm{mm}$. The rods had a length of $133 \,\mathrm{mm}$.
		For the single dark red triangle the diameter of the volume was changed to $200 \,\mathrm{mm}$.}
		\label{fig:Curvature4InfiniteRods}
	\end{figure}

	In the following, we discuss a rectangular configuration of several infinitely long, parallel metallic rods, extending in y direction \cite{hor1998}, which allows to create electric fields in an arbitrary direction in the x-z plane (see Fig.~\ref{fig:Curvature4InfiniteRods}).
	The rods are rectangularly arranged in the x-z plane with a horizontal and vertical spacing of $2 l_{1}$ and $2 l_{2}$, respectively. In the simulation, we apply potentials of $\pm V$ to the four rods as shown in Fig.~\ref{fig:Curvature4InfiniteRods}.

	In Fig.~\ref{fig:Curvature4InfiniteRods}, we show the ratio of curvature to magnitude, $E_{2,v}/E_{0,v}$, of a vertical electric field at the symmetry center of the electrodes as a function of the geometry of the four rod configuration parametrized by $l_1$ and $l_2$.
	To compare different electrode spacings we renormalized the length scale by ${w_{0}}$.
	The dashed line is the corresponding ratio $E_{2,h}/E_{0,h}$ for horizontal fields.
	
	Analytic solutions presented in Fig.~\ref{fig:Curvature4InfiniteRods} clearly show that there is no possible ratio $l_{1} / l_{2}$ at which the absolute values of the two curvatures of the vertical and horizontal field, respectively, are simultaneously minimized. 
	At $l_{1}/l_{2} = 0.58$ the curvature of the vertical field crosses zero, the same happens for horizontal fields at $l_{1}/l_{2} = 1/0.58 = 1.7$. The curvature of an electric field pointing in the vertical direction can thus be canceled when spacing the electrodes by $l_{1}/l_{2} = 0.58$. Still, for horizontal fields the same configuration results in a curvature to amplitude ratio of $E_{2,h}/E_{0,h}= -1.50$.
	
	In the symmetric case, $l_{1} = l_{2} = l$, the two ratios are equal in magnitude but opposite in sign $E_{2,v}/E_{0,v} = - E_{2,h}/E_{0,h} = 0.5 $. For $l= 11~\mathrm{mm}$ we expect a curvature-to-amplitude ratio of $\left| E_{2}/E_{0}\right| = 0.41 ~\mathrm{\frac{1}{cm^2}}$. 
	At $E_{0} = 10~\mathrm{kV/cm}$ this is very close to the limit given in Tab.~\ref{Tab:Limits} and therefore excludes the possibility to use a simple four rod system in our experiment.
	
	Note that the above calculation has been done using analytical methods. When going on to more complicated electrode geometries this is no longer possible. We therefore use the four rod geometry as a benchmark system to estimate possible errors arising from a numerical calculation relying on a finite volume system. This is summarized in Fig.~\ref{fig:Curvature4InfiniteRods}\,B, which shows a comparison between the analytical solution for the four rod system and the numerical simulations performed with COMSOL Multiphysics. As expected, with increasing height and consequently volume of the space in which the electric field is calculated, the values are converging to the analytical solution. For the next section, the volume has been adapted according to this result.

\subsection{Final configuration of eight rods}

\begin{figure}[t]
		\centering
		\includegraphics[width=15cm]{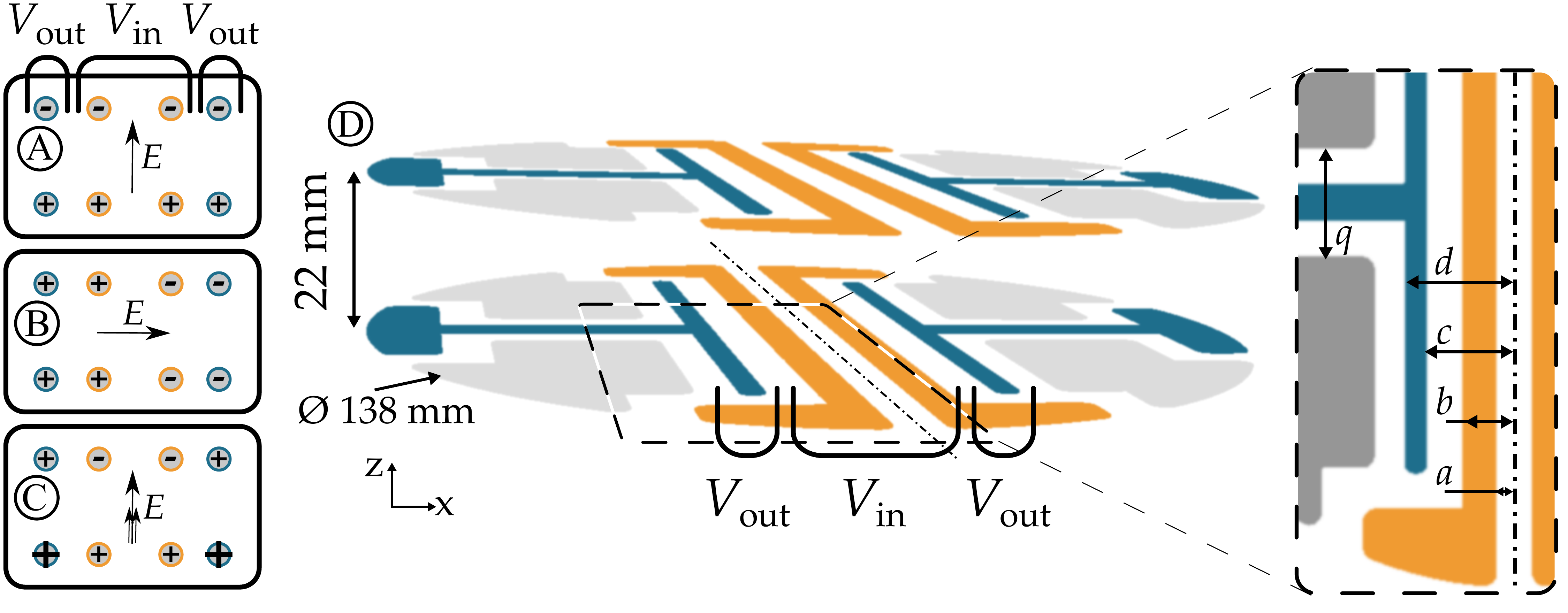}
		\caption{
		Final layout of the ITO electrodes (D).
		The inner (outer) electrodes are colored yellow (blue), the corner patches gray.
		The insets (A), (B) and (C) are cross sections through rod-like electrodes, like in Fig.~\ref{fig:ExpSetup}.
		The outer (inner) circles correspond to the outer (inner) electrodes.
		Three examples how the electrodes can be charged are shown.
		(A): for a uniform field in the z direction. (B): for a uniform field in the x direction is applied. (C): configuration for vertical field as in (A) with a superimposed gradient in z direction.
		(D): the separation of the electrodes on the upper and lower vacuum window is $22 ~\mathrm{mm}$, their diameter $138 ~\mathrm{mm}$.
		The zoomed region of the electrode on the right shows the distances used to optimized the layout.}
		\label{fig:ElectrodesGeometry}
	\end{figure}

	In the previous section we have shown that a four rod system is not sufficient to cancel the curvature of both horizontal and vertical fields.
	A combination of four `inner' and four `outer' electrodes can, however, mutually compensate curvatures.
	If the spacing ratios of inner and outer electrodes fulfill $l_{1}/l_{2} < 0.58$ and $l_{1}/l_{2} > 1/ 0.58$, respectively, the curvatures for vertical and horizontal fields can be canceled. The insets A, B and C of Fig.~\ref{fig:ElectrodesGeometry} illustrate how the desired fields are created. In Fig.~\ref{fig:ElectrodesGeometry}\,A a vertical field is created by applying a positive electric potential to the lower and a negative to the upper electrodes.
	For horizontal fields, as shown in Fig.~\ref{fig:ElectrodesGeometry}\,B, the roles of inner and outer electrodes are exchanged.
	The outer electrodes' spacing ratio is $l_{2}/l_{1} < 0.58$ and its curvature can be compensated by the inner electrodes with $l_{2}/l_{1} > 0.58$.
	Gradients in the vertical direction, as shown in inset C, are obtained by adding an additional potential $+V_{\mathrm{grad}}$ to the outer electrodes.
	The superposition principle allows to create arbitrary field magnitudes and turning angles as linear combinations of the shown potentials.
	
	The implementation of the eight rods, as finally realized on the two opposing main windows, is shown in Fig.~\ref{fig:ElectrodesGeometry}\,D.
	The upper and lower window each carry eight ITO areas. 
	A color code indicates how the most central of those areas correspond to inner and outer rods.
	The four corner areas of the coating, colored light gray area, can be used to compensate unwanted gradients.
	
	The free parameters $a,b,c,d,q$, shown in the detailed view of Fig.~\ref{fig:ElectrodesGeometry} define the geometry.
	They have to be determined by numerical simulations considering one constraint on the distance between any two electrodes. We have to maintain a minimal distance between neighboring electrodes on the same surface of at least 7~mm at assumed potential differences of 10~kV to avoid discharging issues. The minimal distance has been determined by test experiments on air. Discharging properties depend on surface conditioning, but a distance of $7~\mathrm{mm}$ represents a conservative value, which we decided to maintain for the final setup. The parameter $a$ is therefore restricted to $a \geq 3.5~\mathrm{mm}$. 
	Note that the central gap in the ITO coating, described by $a$, also facilitates the use of infrared lasers in the vertical direction.
	
	With the dimension $a$ fixed, the most crucial distance for the overall performance is $b$.
	As the electrodes' mirror symmetry excludes gradients in the electric field magnitude, the major goal is to minimize the curvature-to-amplitude ratio $E_{2}/E_{0}$ for vertical as well as horizontal electric fields. 
	
	For the simulations of the electric field, we imported the electrode designs with variable parameters $b,c,d$ and $q$ in COMSOL Multiphysics and the magnitudes of the electric fields generated by the outer and the inner electrodes were computed independently. 
	The direction of the electric field generated by inner and outer electrodes only slightly differs over the assumed dipole trap volume and this justifies to use the electric field magnitudes instead of the electric field vectors. Up to $w_{0} = 0.1 ~\mathrm{mm}$ away from the symmetry center vertical fields created by the inner or the outer electrodes deviate by less than $10^{-4}$ from the $z$ direction. These small deviations of the direction enter in the field magnitude quadratically and result in an error below $10^{-8}$.
	
		\begin{figure}[b]
		\centering
		\includegraphics[width=12cm]{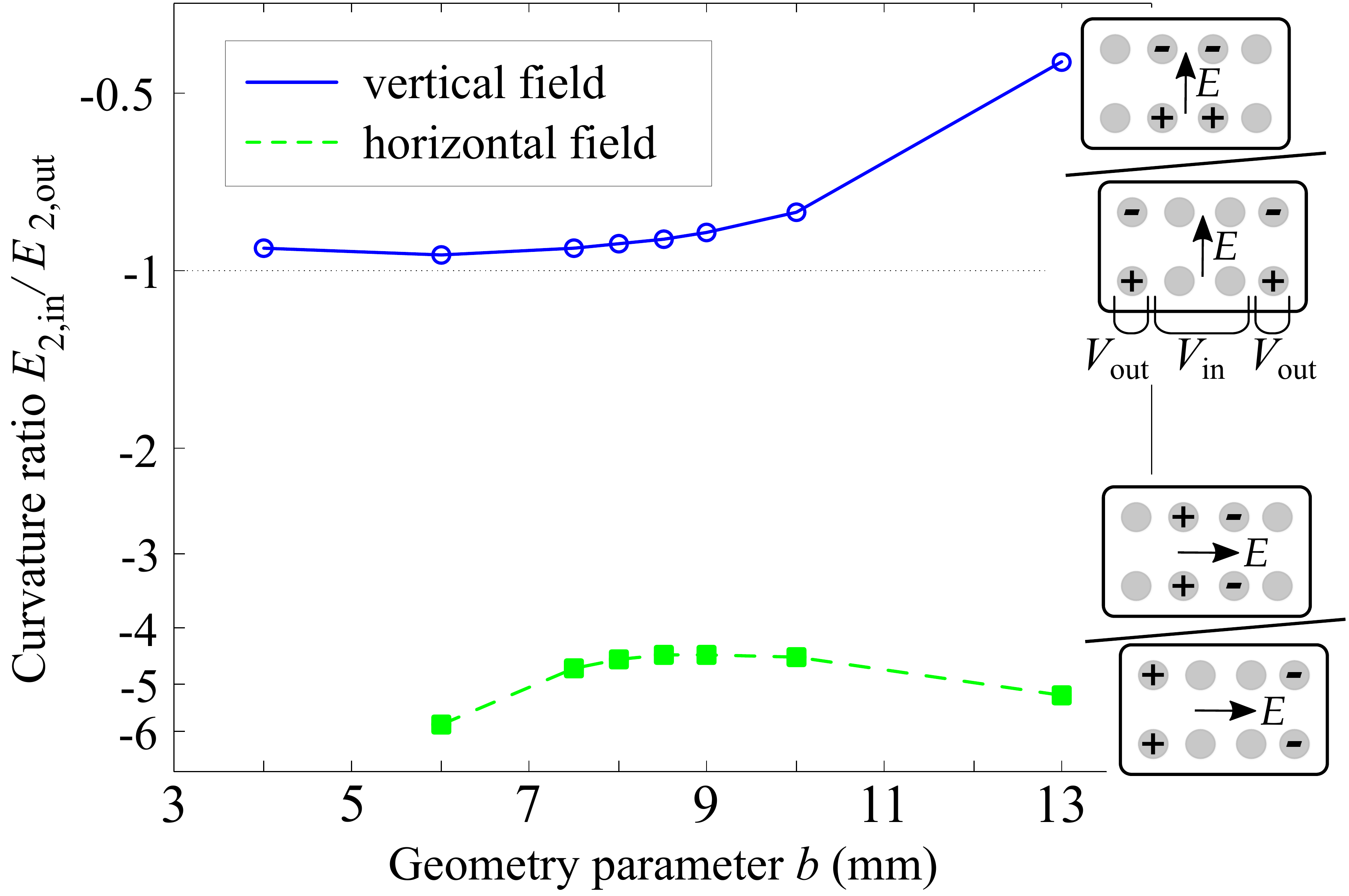}
			\caption{
			Curvature ratio $E_{\mathrm{2,\mathrm{in}}}/E_{\mathrm{2,\mathrm{out}}}$ as a function of $b$.
			The data for the curvature ratios of vertical and horizontal fields are represented by empty circles and full squares respectively.
			The dotted line indicates the desired value of $-1$ for vertical fields.
			The insets indicate for which configurations the ratios where taken.}
		\label{fig:AmplAndCurvRatios}
		\end{figure}
	
	In a first step, we consider the eight rod geometry with applied potentials as indicated in Fig.\,\ref{fig:AmplAndCurvRatios} for the special case where $|V_{\mathrm{in}}|$=$|V_{\mathrm{out}}|$. The goal is to optimize parameter $b$ for minimal total curvature $E_2=E_{2,\mathrm{in}}+E_{2,\mathrm{out}}$, ideally $E_2=0$, which is equivalent to a ratio $E_{2,\mathrm{in}}/E_{2,\mathrm{out}}=-1$. Figure \ref{fig:AmplAndCurvRatios} shows $E_{2,\mathrm{in}}/E_{2,\mathrm{out}}$ as a function of $b$ for the vertical and horizontal field, respectively. For the vertical field, the ratio $E_{2,\mathrm{in}}/E_{2,\mathrm{out}}$ is close to $-1$ over a large range of values of $b$.	Only for very large $b$ the curvature ratio is going to zero, as the inner electrodes extend far in the region with a ratio $l_{1}/l_{2} > 0.58$ and compensate their field curvature themselves.
	When switching the potentials to obtain horizontal fields, inner and outer electrodes change roles and we obtain a ratio $E_{2,\mathrm{in}}/E_{2,\mathrm{out}}$ as shown by the green line in Fig.\,\ref{fig:AmplAndCurvRatios}. $|E_{2,\mathrm{in}}/E_{2,\mathrm{out}}|$ is minimal at $b\approx 9\,\mathrm{mm}$ and we fix $b$ to this value.
	
	To finally cancel the curvature for both vertical and horizontal fields in the eight rod system, the potential ratio $V_{\mathrm{in}}/V_{\mathrm{out}}$ of the electrode systems has to be tuned to the inverse of the curvature ratio $E_{\mathrm{2,in}}/E_{\mathrm{2,out}}$.
	For $b = 9~\mathrm{mm}$ curvature can be exactly canceled by applying the potentials with a ratio $V_{\mathrm{in}}/V_{\mathrm{out}} = 1/0.86$ for vertical fields and $V_{\mathrm{in}}/V_{\mathrm{out}} = 1/4.4$ for horizontal fields.
	Note that, in the case of vertical electric fields, the inner electrode contribution to the magnitude of the electric field is 11 times larger when compared to the contribution of the outer ones, which are mainly acting to cancel the curvature generated by the inner ones.
	
	The same analysis can be performed for the other geometrical parameters.
	As the desired spacing between electrodes is $7~\mathrm{mm}$, $c$ cannot become smaller than $c=b+7~\mathrm{mm}=16.0 ~ \mathrm{mm}$. 
	Furthermore, we observe that the results depend only slightly on the width of the outer electrode determined by $d$.
	To enhance the effect of the corner patches we have consequently chosen $d$ and also $q$ small.
	The main goal of the corner patches is to compensate unforeseen imperfections in manufacturing like a relative tilt of the vacuum windows. The patches are capable of counterbalance a gradient of $2.2 \cdot 10^{-2} ~ \mathrm{kV/cm^2}$ when the molecular sample is polarized to $\mathrm{68\%}$.
	This gradient would appear for a vacuum window tilt of about $0.29^\circ$, which is far larger than the measured one of $\left( 0.024 \pm 0.042 \right)^\circ$.
	
	In summary the parameters of the electrodes presently installed on our vacuum chamber are: $a = 3.5 ~ \mathrm{mm}$, $b = 9.0 ~ \mathrm{mm}$, $c = 16.0 ~ \mathrm{mm}$, $d = 19.0 ~ \mathrm{mm}$, $q = 20.0 ~ \mathrm{mm}$.
	A gradient allowing to address transitions between rotational states in single layers as considered in Fig.~\ref{ElectricalFieldGradientsInOpticalLattices} can be applied by adding a potential $V_{\mathrm{grad}}$ to the outer electrodes.
	For example, by adding $V_{\mathrm{grad}} = +8.46~\mathrm{kV}$ to all outer electrodes results in a gradient of $3 ~\mathrm{kV/cm^2}$.

\section{Conclusion} 

	In this paper, we discussed and presented a sophisticated electrode system for the manipulation and control of ultracold polar NaK molecules.  Particular care has been taken to realize largely homogeneous fields with low curvature to keep perturbations of the optical trapping potentials as low as possible. Furthermore, the final electrode configuration allows to realize strong electric field gradients across an optical lattice for single site addressing of molecules. The final eight rod geometry is realized by a thin ITO coating on the main vacuum windows to allow for large optical access and the realization of versatile optical trapping potentials for the simulation of dipolar quantum many-body physics.
	
	 In the future, the designed ITO based electrode geometry could be scaled down to  smaller sized electrodes and could potentially be realized in an `atom-chip' like design  \cite{sch2006, ste2014}. 
	 An integrated design by means of ITO coated atom chips would allow for complete optical access to the molecules even from below the surface of the chip. Due to the small length scales of the chip, particular care would have to be taken to cancel gradients, but  required electric potentials would be lower by several orders of magnitudes allowing for fast and versatile control of the electric fields.  
	
\section*{Acknowledgements} 
	We acknowledge financial support from the Centre for Quantum Engineering and Space-Time Research QUEST and the European Research Council through ERC Starting Grant POLAR.
	M.G. and T.H. acknowledge the support from the Research Training Group 1729,
	K.V. from the Research Training Group 1991.
	The deposition of the optical layer system and the indium tin oxide electrodes were performed at the Fraunhofer IST in Braunschweig, Germany with the optical coating system $\mathrm{EOSS\textregistered}$ by Daniel Rademacher, Stefan Bruns, Tobias Zickenrott, and Thomas Neubert.
	
\section*{References}

\bibliographystyle{iopart-num}

\providecommand{\newblock}{}

\end{document}